\documentclass[preprint,showpacs,preprintnumbers,amsmath,amssymb]{revtex4}

\usepackage{graphicx}
\usepackage{dcolumn}
\usepackage{bm}
\usepackage{color}

\newcommand{\be}{\begin{eqnarray}}
\newcommand{\ee}{\end{eqnarray}}
\renewcommand{\vr}{\mathbf{r}}
\newcommand{\vu}{\mathbf{u}}
\newcommand{\vro}{\mbox{\boldmath$\rho$}}
\newcommand{\vX}{\mathbf{X}}
\newcommand{\vze}{\mathbf{0}}

\newcommand{\ds}{\displaystyle}

\newcommand{\wh}{\widehat}

\begin{document}

\title{Kinetics of diffusion-limited catalytically-activated reactions:
An extension of the Wilemski-Fixman approach}

\author{O. B\'enichou}
\email{benichou@lptl.jussieu.fr} \affiliation{Laboratoire de
Physique Th\'eorique de la Mati\`ere Condens\'ee, 4 place Jussieu,
75252 Paris Cedex 05}
\author{M. Coppey}
\email{coppey@lptl.jussieu.fr} \affiliation{Laboratoire de Physique
Th\'eorique de la Mati\`ere Condens\'ee, 4 place Jussieu, 75252
Paris Cedex 05}
\author{M. Moreau}
\email{moreau@lptl.jussieu.fr} \affiliation{Laboratoire de Physique
Th\'eorique de la Mati\`ere Condens\'ee, 4 place Jussieu, 75252
Paris Cedex 05}
\author{G. Oshanin}
\email{oshanin@lptl.jussieu.fr} \affiliation{Laboratoire de Physique
Th\'eorique de la Mati\`ere Condens\'ee, 4 place
Jussieu, 75252 Paris Cedex 05}
\affiliation{Max-Planck-Institut f\"ur Metallforschung,
Heisenbergstr. 3, D-70569 Stuttgart, Germany} \affiliation{Institut
f\"ur Theoretische und Angewandte Physik,University of Stuttgart,
Pfaffenwaldring 57, D-70569 Stuttgart, Germany}

\begin{abstract}
We study kinetics of diffusion-limited catalytically-activated $A +
B \to B$ reactions taking place in three dimensional systems, in
which an annihilation of diffusive $A$ particles by diffusive traps
$B$ may happen only if the encounter of an $A$ with any of the $B$s
happens within a special catalytic subvolumen, these subvolumens
being immobile and uniformly distributed within the reaction bath.
Suitably extending the classical approach of Wilemski and Fixman (G. Wilemski and M. Fixman, J. Chem. Phys. \textbf{58}:4009, 1973) to
such three-molecular diffusion-limited reactions, we calculate
analytically an effective reaction constant and show that it
comprises several terms associated with the residence and joint
residence times of Brownian paths in finite domains. The effective
reaction constant exhibits a non-trivial dependence on the reaction
radii, the mean density of catalytic subvolumens and particles'
diffusion coefficients. Finally, we discuss  the fluctuation-induced
kinetic behavior in such systems.
\end{abstract}

\pacs{05.40.-a, 02.50Ey, 82.20.-w}
\maketitle

\section{Introduction}

Catalytically-activated reactions involving diffusive species
underly many different processes in physics, chemistry and biology
\cite{bond,rice,moreau95,oshanin98,tox}. In a general notation, such
reactions can be written as
\begin{equation}
\label{rea} A+B+C \xrightarrow[]{k}  P+C,
\end{equation}
where $A$ and $B$ designate two different types of mobile reactive
species, $C$ denotes a catalytic subvolumen, while $P$ stands for
the reaction product. Catalytic subvolumens $C$ may form some
patterns, be uniformly or regularly spread along a given structure,
(e.g., a polymer or polymers in solution), scattered uniformly
within or on the boundary (surface) of the reaction bath.

The reaction scheme in Eq.(\ref{rea}) signifies that the bimolecular
reaction between the $A$ and $B$ molecules may take place, at some
finite elementary reaction rate $k$, if and only if a diffusive
encounter of an
 $A$ and a $B$ happens within any catalytic subvolumen $C$. In some cases, the $B$ particles may be unaltered by the reaction or
their concentration may substantially exceed that of the $A$ species, in
which situation the reaction in Eq.(\ref{rea}) can be viewed as \textit{bi-catalytic}:
 that is, one deals with a simplified reaction scheme $A\to
P$ which requires  the presence of two different catalytic
subvolumens - $B$ and $C$.
This is most often the case in biology, as exemplified, for
instance,
 by the transcription of genes induced by the simultaneous presence of
several transcription factors on the promoter sequence \cite{biotf}.

Most of analytical descriptions of the catalytically-activated
reactions have focussed so far on the particular question how
reactions are promoted by specific catalytic subvolumens, which was
believed to be the most crucial aspect of the problem \cite{feib}. Within this
line of thought, the kinetic behavior has been determined using
standard formal-kinetic approaches \cite{bond,rice}. On the other
hand, a few available analytical studies of the
catalytically-activated reactions limited both by diffusion of
species and by the condition of the simultaneous encounters within
the catalytic subvolumens have revealed a non-trivial kinetic
behavior in low dimensional systems, and showed that although in
three dimensions kinetics follows standard temporal behavior, the
effective reaction rates are strongly dependent on particles'
diffusion coefficients, subvolumens' radii and their concentration
\cite{moreau95,oshanin98,tox}. These findings are, of course, in an
apparent contradiction with the predictions of the formal-kinetic
approach.

As a matter of fact, it has been already realized that for reactions
taking place in non-catalytic systems, in many instances, the
kinetic behavior can not be adequately  described in terms of the
formal-kinetics approach. Indeed, it has been known for a long time
that diffusion of reactive species limits the reactive process and
leads to unusual kinetics in low dimensional systems (see, e.g.,
Ref.\cite{klafter} and references therein). Moreover, it has been
discovered that in many reactive
 systems spatial fluctuations in particles concentrations dominate the
 long-time evolution and entail anomalous, fluctuation-induced
 behavior. In particular,
 a pronounced deviation from the
conventional descriptions \cite{rice} has been predicted for the
irreversible, diffusion-controlled recombination reaction $A + B \to
0$ in case when initially the particles of the $A$ and $B$ species
are all distributed at random, independently of each other and with
strictly equal mean densities $n_A(0) = n_B(0) = n_0$. It has been
first shown \cite{ov} that here at long times the mean particle
densities follow
\begin{equation}
\label{AB} n(t) \sim n^{1/2}_0 (D t)^{-d/4},
\end{equation}
where $d$ is the space dimensionality and $D$ - the sum of
particles' diffusion coefficients, $D = D_A + D_B$. This law, which
was rigorously proven in Refs.\cite{burl,leb}, should be contrasted
to the conventionally expected Smoluchowski-type form $n(t) \sim
1/\phi_R^{(d)}(t)$ \cite{rice}, where, as $t \to \infty$,
\begin{equation}
\label{Smol} \phi_R^{(d)}(t) = \int_0^t d\tau K_S(\tau) \sim
\displaystyle \left\{\begin{array}{lll}
\displaystyle 4 \sqrt{D t/\pi},  \;\;\;   \mbox{d = 1}, \nonumber\\
\displaystyle \frac{4 \pi D t}{\ln(4 D t/R^2)},  \;\;\;   \mbox{d = 2}, \nonumber\\
\displaystyle 4 \pi D R t,    \;\;\;  (k = \infty) \;\;\;   \mbox{d = 3},
\end{array}
\right.
\end{equation}
$K_S(\tau)$ being the $d$-dimensional Smoluchowski-type constant,
defined as the flux of diffusive particles through the surface of an
immobile sphere of radius $R$ - the reaction radius. Note that both
decay laws contradict to the text-book formal-kinetic description based on the
"law of mass
action", which predicts that regardless of the spatial
dimensionality
 $n(t)$ decays as $n(t) \sim 1/t$ \cite{rice}.

 Therefore, according
 to the Smoluchowski approach, in diffusion-controlled recombination reaction $A + B \to
0$ diffusion slows down the decay in low
 dimensional systems and entails the renormalization of the reaction
 rates in three-dimensions. On the other hand, in the particular case
 when initially the particles of the $A$ and $B$ species
are all distributed at random, independently of each other and with
strictly equal mean densities $n_A(0) = n_B(0) = n_0$, fluctuations
in
 spatial distributions of the reactive species appear as the most important rate
 controlling factor in spatial dimension $d \leq 4$ and dominate the
 long-time kinetics.

For trapping reactions $A + B \to B$ in completely catalytic systems
two opposite limiting situations were most thoroughly studied.
 Namely, the case
when the A particles diffuse while the traps $B$ are static, and the
situation in which the $A$ particles are immobile while $B$s diffuse
- the so-called target annihilation problem (TAP). The case of
static traps has attracted most of interest prompted by, in part,
 an early
observation \cite{bal} that the long-time survival probability
$P_A(t)$ of $A$ particles diffusing in the presence of randomly
placed (with mean density $[B]$) traps exhibits highly non-trivial,
fluctuation-induced behavior of the form
\begin{equation}
\label{traps} \ln P_A(t) \sim - [B]^{2/(d + 2)} (D_A t)^{d/(d+2)},
\;\;\; t \to \infty,
\end{equation}
which stems from the randomness of $B$ distribution in space and
namely, from the presence of large spatial regions devoid of traps
where the $A$ particles survive anomalously long times. This
fluctuation-induced decay law is intrinsically relevant to the
so-called Lifschitz singularities near the edge of the band in the
density of states of a particle in quantum Lorentz gas, as first
noticed in \cite{bal}. Later works (see, e.g., Refs.\cite{burl,don,bal,3,sosiska,pastur,leb,gp,kh,weiss,fix,deutch,mi})
have also pointed out relevance of the issue  to the problems of
percolation, self-avoiding random walks or self-attracting polymers,
as well as anomalous behavior of the ground-state energy of the
Witten's toy Hamiltonian of supersymmetric quantum mechanics
\cite{sosiska}.

Survival probability $P_{target}(t)$ of an immobile target $A$ of
radius $R$ in presence of point-like diffusive traps $B$ - the
 target annihilation problem (TAP), allows for an exact
solution in any spatial dimension
\cite{klafter,tach,blu,red2,szabo,burl}
\begin{equation}
\label{k} P_{target}(t) = \exp\Big( - [B] \phi_R^{(d)}(t)\Big),
\end{equation}
where $\phi_R^{(d)}(t)$ has been defined in Eq.(\ref{Smol}) in which
one has to set $D_A = 0$. Extensions to systems with hard-core
interactions between traps \cite{core} or fluctuating chemical
activities of traps \cite{fluct} have been also provided.

Now, the general and physically most important case of trapping
reactions when both $A$ and $B$s diffuse with diffusion coefficients
$D_A$ and $D_B$ was not solved exactly up to the present time. It
has been proven \cite{leb} that in this case the $A$ particle
survival probability obeys
\begin{equation}
\label{general} \ln P_A(t) = - \lambda_d(D_A,D_B) \times
\displaystyle \left\{\begin{array}{lll}
\displaystyle t^{1/2},  \;\;\;   \mbox{d = 1}, \\
\displaystyle \frac{t}{\ln(t)},  \;\;\;   \mbox{d = 2}, \\
\displaystyle t,    \;\;\;    \mbox{d = 3},
\end{array}
\right.
\end{equation}
which defines its time-dependence exactly.

On the other hand, the factor $\lambda_d(D_A,D_B)$ remained for a
long time an unknown function of the particles' diffusion
coefficients and spatial dimension. Only very recently some rigorous
arguments have been presented showing that  for  $A+B\to B$
reactions
 in low dimensions \cite{bly,benbis,mor,blybis},  the long time decay of $A$
particles concentration is exactly as in the TAP problem,
Eq.(\ref{k}), and thus is asymptotically \textit{independent} of the
$A$ particles diffusion coefficient even  in the case when both
species diffuse.

In three dimensions, however, the precise form  of
$\lambda_d(D_A,D_B)$ is still undetermined and still very little is
known about it. It has been shown that $\lambda_d(D_A,D_B)$ is less
than the rate constant calculated within the Smoluchowski approach
\cite{burl} and moreover, it has been realized
that in case when $D_A$ and $D_B$ are
sufficiently small, $\lambda_d(D_A,D_B)$ may be bounded by a
non-analytic function of particles' diffusion coefficients
\cite{bere}. A perturbation theory approach for calculation of
$\lambda_d(D_A,D_B)$ has been proposed in Ref.\cite{szabo} and the
corrections to
 the predictions of the Smoluchowski approach have been evaluated.
It was also shown that $\lambda_d(D_A,D_B)$ can not be represented
as the function of the combination $D = D_A + D_B$ only, since the
diffusion-reaction equations are not separable. Therefore, even in completely catalytic systems
the evaluation of $\lambda_d(D_A,D_B)$ represents a fairly
complicated many-body problem.

In this paper we study the kinetics of the catalytically-activated
diffusion-limited reactions in Eq.(\ref{rea}) in the special case
when $B$ particles remain unaltered by reactions, i.e. the case of
diffusion-limited catalytically-activated trapping reactions,
description of which poses such serious technical difficulties even
in  the non-catalytic systems (see the discussion above). In order
to obtain an effective reaction rate for such bi-catalytic reactions
taking place in a
 homogeneous three-dimensional medium, we first develop an analytical
approach, inspired by the work of Wilemski and Fixman
\cite{wilemski73}, which allows one to estimate the reaction rate
for non-catalytic bimolecular reactions. Here we extend this
Wilemski-Fixman approach (WFA) to catalytically-activated trapping reactions.

We assume that the catalytic subvolumens $C$ are
 immobile and are spread uniformly in the reaction
bath. On the other hand,  $A$ and $B$ particles are assumed to
perform unconstrained diffusion, and react only when they are
simultaneously present in a spherical  domain of radius $R$ centered
around each catalytic subvolumen $C$. As well, we suppose that there
is no other interaction between the particles except for the
reaction, which enables us to describe the $A$ particle dynamics in
terms of a Fokker-Planck-type equation with a sink term which mimics
the presence of the traps $B$ and of the catalytic subvolumens $C$.
In order to obtain a closed equation, we follow then the well known
Wilemski-Fixman approach (WFA) \cite{wilemski73}. This approximation
relies on the time and space separation of the joint probability
density; that is, the probability density is assumed to be a product
of the equilibrium density and of a certain time-dependent function,
supposing that initially the system is at equilibrium. We hasten to
remark, that, although the validity domain of this approximation is still not really known, many researchers have shown that the WFA describes
quite correctly reaction kinetics in several general situations
\cite{doi75,weiss1}. For example, Do\"\i{} \cite{doi75} showed that
the WFA can be used for the purely diffusion-limited case ($k =
\infty$), contrarily to the intuitive expectation that the WFA is
appropriate for systems with a weak chemical reaction rate $k$. We
will show also in what follows that the effective reaction constant
obtained within such an approach reduces to the well-established
results in several limiting cases.

Finally, we will present an estimate of the impact of the
fluctuation effects on the long-time kinetics of the
diffusion-controlled catalytically-activated reactions.

The paper is outlined as follows: In section II, we formulate the
model and write down basic equations.  Section III is devoted to the
solution of the evolution equations within the framework of the
suitably extended Wilemski and Fixman approach. Here, we determine
an effective reaction rate for the bi-catalytic reaction and show
that it can be is expressed through different functionals of
Brownian motion, known as \textit{residence times} and \textit{joint
residence times} of Brownian particles in some specified domains.
Further on, in Section IV we calculate the residence times involved
and obtain an explicit expression for the effective rate constant
describing diffusion-limited catalytically-activated trapping
reactions. As well, we discuss its asymptotical behavior in several
limiting situations. Next, in Section V we present some estimates of
the long-time fluctuation-induced behavior in such systems. Finally,
in Section VI we conclude with a brief summary of our results and
discussion.

\section{Model and basic equations}

\begin{figure}[htb]
\begin{center}
\includegraphics[scale=.5]{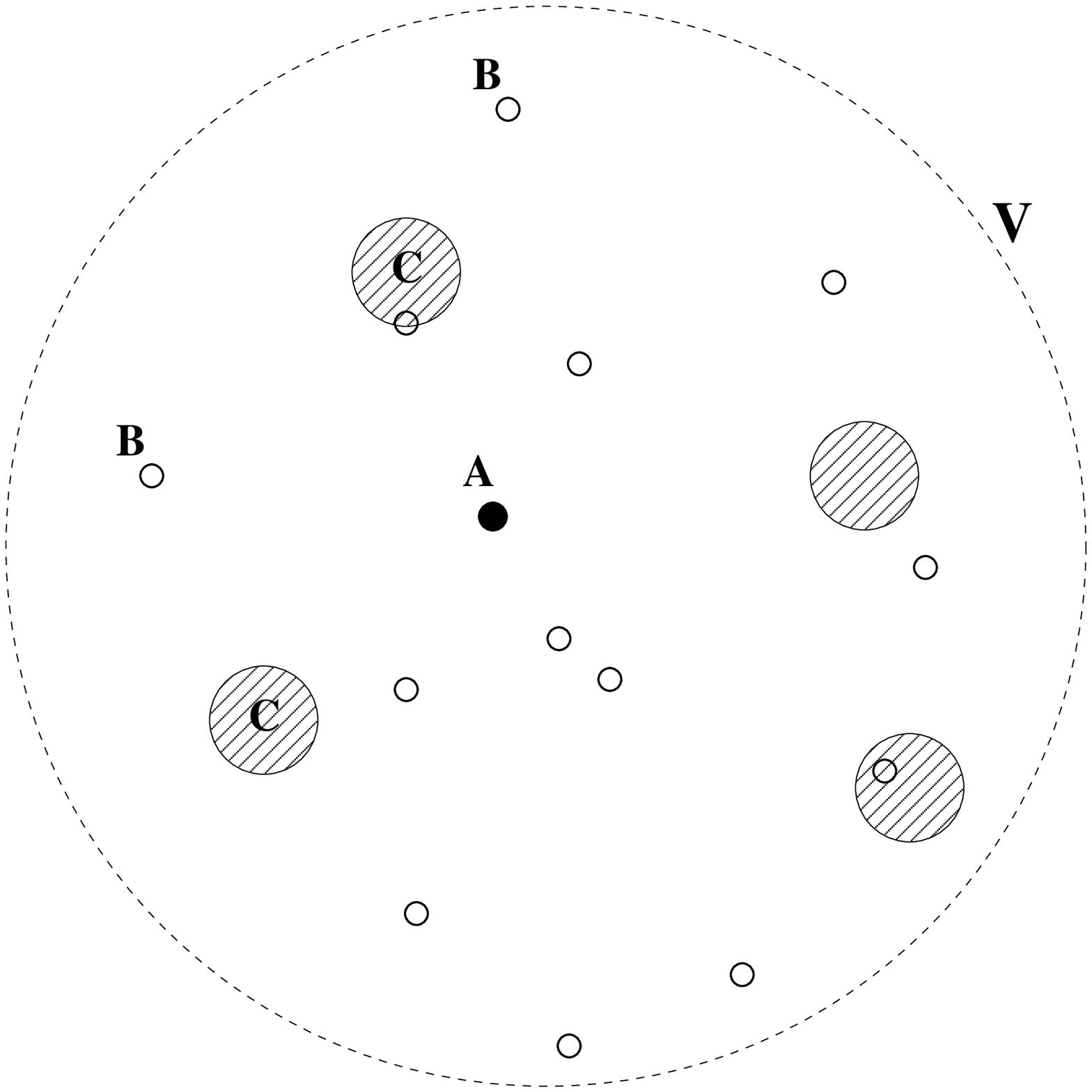}
\end{center}
\caption{\label{f1}}
\end{figure}

\begin{figure}[htb]
\begin{center}
\includegraphics[scale=.6]{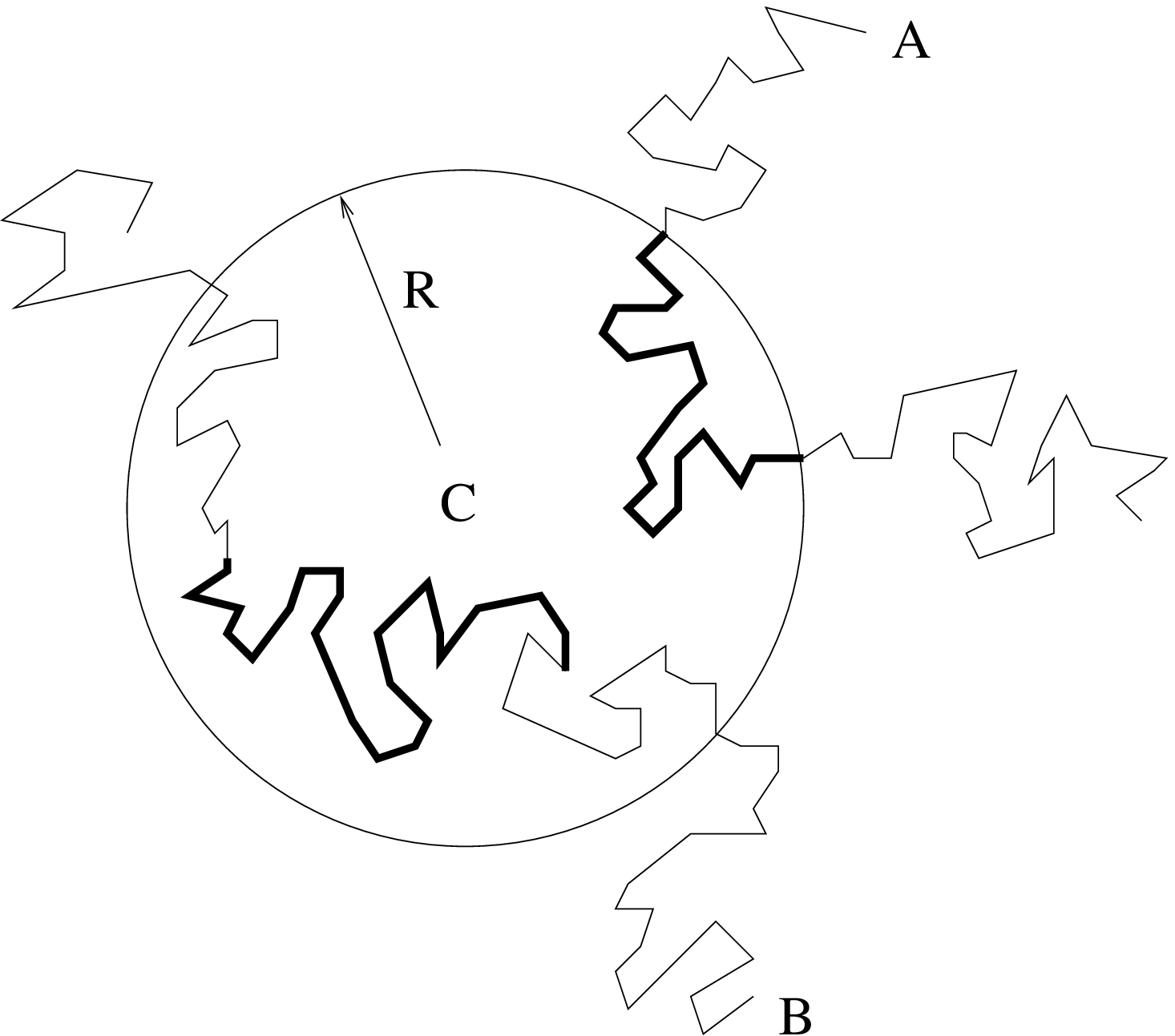}
\end{center}
\caption{\label{figtps1}}
\end{figure}

Consider a three-dimensional reaction bath of volume $V$ comprising
a single $A$ particle,  $m$ traps $B$, and $q$ immobile catalytic
subvolumens $C$ {(see Fig.\ref{f1})}. These catalytic subvolumens
are uniformly distributed in the reaction volume with mean
concentration $[C]$. The $A$ and $B$ particles diffuse freely with
diffusion coefficients $D_A$ and $D_B$, respectively. For
simplicity, we assume that $A$ and $B$s are point-like particles of
zero radius such that excluded-volume interactions between them can
be safely neglected. Now, the reaction between particles - an
annihilation of the $A$ particle by any of the $B$s - takes place
with a given probability defining the elementary reaction constant
$k$ when both species appear simultaneously within a spherical
region of radius $R$ (which can be also thought off as the reaction
radius) centered around any catalytic subvolumen $C$. Such a
"reactive" situation is depicted in Fig.\ref{figtps1}.

We now proceed by suitably extending the celebrated approach devised originally by Wilemski and Fixman \cite{wilemski73} for non-catalytic trapping $A + B \to B$
reactions. The basic idea behind this approach is that the presence of traps $B$ can be effectively modelled by introducing a sink function $S$
into the diffusion equation describing dynamics of the $A$ and $B$ particles. Then, the sink term determines the efficiency of the reaction as a function of the instantaneous separation distance between the $A$ particle and the traps $B$.

In the simplest formulation, this sink term can be represented as the Heaviside function, which implies that the reaction takes place with an
elementary reaction constant
 $k$
as soon as the $A$ particle appears in the vicinity of any of the traps $B$. Adapting this line of thought, we describe the $A$
and $B$ particles dynamics in terms of the following multivarient diffusion equation:

\be \label{fp} \frac{\partial \Psi}{\partial t}= D_A \nabla_A^2 \Psi + D_B \sum_i^m \nabla_{B_i}^2\Psi - k S\psi \ee
where
 $\Psi(\{\vr\},t)$ is time-dependent $m+1$ particles probability density
 function, $\{\vr\}=\{\vr_A,\vr_{B_1},...,\vr_{B_m}\}$ defines the positions
of the $A$ particle and all $m$ traps $B$, while the sink function
$S$ is represented as a set of Heaviside functions $H(x)$, ($H(x)=1$
if $x\geq 0$ and $H(x)=0$ for $x<0$), centered around $q$ catalytic
subvolumens:

\be \label{Sdef} \nonumber S&=&\frac{1}{V'^2} \sum_{i=1}^m \sum_{k=1}^q
H(R-|\vr_A-\vr_{C_k}|) H(R-|\vr_{B_i}-\vr_{C_k}|) \\
&\equiv&\frac{1}{V'^2} \sum_{i=1}^m \sum_{k=1}^q
H_{A,C_k}H_{B_i,C_k}, \ee in which
 $V'$ denotes the volume of a "reactive" domain, $\ds V'=\frac{4}{3}
\pi R^3$, (see Fig.\ref{figtps1}).

Consequently, the desired probability $P(t)$ that the $A$ particle survives up to time $t$ obeys:

\be \label{psdef} P(t)=\int d\{\vr\} \Psi(\{\vr\},t), \ee
and can be readily evaluated once $\Psi(\{\vr\},t)$ is known.

\section{Solution of the evolution equation.}

Equation (\ref{fp}) can be cast into an equivalent form by
 using the Green function $G(\{\vr\},t;\{\vr^0\},t^0)$ of equation
(\ref{fp}) without the sink term, the latter being considered as an inhomogeneity \cite{wilemski73}. In doing so, we find that
 the formal solution of equation (\ref{fp}) reads:

\be \label{q} \nonumber \Psi(\{\vr\},t)=\int d\{\vr'\} \Psi(\{\vr'\},0)
G(\{\vr\},t;\{\vr'\},0)\\
- k\int_0^t dt^0 \int d\{\vr^0\} G(\{\vr\},t;\{\vr^0\},t^0) S(\{\vr^0\}) \Psi(\{\vr^0\},t^0). \ee
Note that $G(\{\vr\},t;\{\vr^0\},t^0)$ of equation (\ref{fp}) without the sink term factorizes:

\be G(\{\vr\},t;\{\vr^0\},t^0)=G_A(\vr_A,t;\vr_A^0,t^0) \prod_{u=1}^mG_B(\vr_{B_u},t;\vr_{B_u}^0,t^0). \ee
Now,  supposing that at $t = 0$ the traps $B$ were uniformly distributed in the reaction bath, i.e. that $\Psi(\{\vr'\},0)=\Psi(0)$, we find that
the first term on the right-hand-side of Eq.(\ref{q}) reduces to $\Psi(0)$.
Following the reasonings of Wilemski and Fixman, we assume further on that $\Psi(0)$ is the equilibrium density $\Psi_{eq}=1/V^{m+1}$. This
corresponds to the physical situation when the system is first brought to equilibrium and the reaction is triggered then at time $t=0$. Next,
multiplying both sides of Eq.(\ref{q}) by $S(\{\vr\})$ and integrating it over all spatial variables $\{\vr\}$, we obtain:

\be \label{vt} \nonumber v(t)=v_{eq}-k\int_0^t dt^0 \int d\{\vr\}\int
d\{\vr^0\} S(\{\vr\})  \\
\times G(\{\vr\},t;\{\vr^0\},t^0)S(\{\vr^0\}) \Psi(\{\vr^0\},t^0)
\ee where, by definition, $v(t)\equiv\int d\{\vr\} S(\{\vr\})
\Psi(\{\vr\},t)$ and, equivalently, $v_{eq}\equiv\int d\{\vr\}
S(\{\vr\}) \Psi_{eq}$. Using next the definition of $S(\{\vr\})$
given in equation (\ref{Sdef}), we obtain for $v_{eq}$:

\be \nonumber v_{eq} & = & \frac{1}{V'^2}\frac{1}{V^{m+1}}\sum_{i=1}^m
\sum_{k=1}^q V^{m-1}\int d\vr_Ad\{\vr_{B_j}\}H_{A,C_k} H_{B_i,C_k} \\
\nonumber & = & \frac{1}{V'^2}\frac{1}{V^{m+1}}\sum_{i=1}^m \sum_{k=1}^q V^{m-1} V'^2
 =  \frac{mq}{V^2}. \ee

Further on, integrating equation (\ref{fp}) over all $\{\vr\}$, and
using the definition of the survival probability (\ref{psdef}), we
find that:

\be \label{relphiv} \frac{d P(t)}{dt}=-kv(t). \ee

Consequently, the function $v(t)$ determines the rate of the time evolution of the $A$ particle survival probability.

In order to obtain a closed equation for $v(t)$, we follow again
Wilemski and Fixman method, assuming that $\Psi(\{\vr\},t)$
can be split into the product of a time-dependent function and the
equilibrium density corresponding to the situation without reaction,
knowing that initially $\Psi(\{\vr\},0)=\Psi_{eq}$. This
approximation is valid \textit{a priori} for a small enough value of
the reaction rate $k$, such that the probability density can be
thought of being close to the equilibrium's one at any time. As a
matter of fact, this approximation still holds in much more general
situations, for example when the reaction is
diffusion-limited ($k = \infty$), as it was shown by Do\"\i{} \cite{doi75}. A
detailed discussion of this approximation can be found in Refs.
\cite{weiss,battezzati,perico}.  Applying it to our
case of the catalytically-activated diffusion-limited trapping
reactions, we have:

\be \Psi(\{\vr\},t)\approx \Psi_{eq} \nu (t). \ee
Noticing next that

\be \nonumber \int d\{\vr\} S \Psi &=& \nu(t) \int d\{\vr\} S
\Psi_{eq}=\nu(t) v_{eq} \equiv v(t), \ee we get $\nu(t)=v(t)/v_{eq}$
and hence, the approximated probability density reads:

\be \Psi=\Psi_{eq} \frac{v(t)}{v_{eq}}. \ee The latter equation,
within the framework of the Wilemski-Fixman approximation, yields
the following result for $v(t)$:

\be \label{vtI} v(t)=v_{eq} -k\frac{\Psi_{eq}}{v_{eq}}\int_0^t dt^0 v(t^0) I, \ee
where the integral $I$ is given explicitly by:

\be I&=& \frac{1}{V'^4} \sum_{i,j,k,l} \int d\{\vr\}\int d\{\vr^0\} H_{A,C_k} H_{B_i,C_k} H_{A^0,C_l} H_{B_j^0,C_l} G(\{\vr\},t;\{\vr^0\},t^0).
\ee

In order to obtain an explicit expression for this integral, and
thus to get an access to the kinetic behavior of the survival
probability, we consider it in more detail below. We note that $I$
can be split into four parts, when specifying the following different events:
the $A$ particle, initially present in the $k$-th subvolumen $C$
together with the $i$-th particle $B$, will further encounter either
the same $B$ particle ($i=j$) or another $B$ particle ($i\neq j$) in
either the same subvolumen $C$($k=l$) or some other subvolumen $C$
($k\neq l$).

First, let us consider the integral $I_{ijkl}$ for $i\neq j$. Integrating over $\vr_{B_u}$ for $u\neq i$ and over $\vr^0_{B_v}$
for $v\neq j$, we obtain:

\be I_{ijkl;i\neq j}&=&\frac{V^{m-2}}{V'^4} \int d\vr_A d\vr_A^0 d\vr_{B_i} d\vr_{B_j}^0 H_{A,C_k} H_{B_i,C_k} H_{A^0,C_l} H_{B_j^0,C_l}
G_A(\vr_A,t;\vr_A^0,t^0). \ee
Next, performing the integration of the Heaviside functions $H_{B_j^0,C_l}$ and $H_{B_i,C_k}$ over  $d\vr_{B_i}$ and $d\vr_{B_j}^0$, we have:

\be I_{ijkl;i\neq j}&=&\frac{V^{m-2}}{V'^2} \int d\vr_A d\vr_A^0 H_{A,C_k} H_{A^0,C_l} G_A(\vr_A,t;\vr_A^0,t^0). \ee

Let us now consider the integral $I_{ijkl}$ for $i=j$. Integrating over the variables which do not appear as arguments of the
Heaviside functions, we can write that:

\be \nonumber I_{iikl}&=&\frac{V^{m-1}}{V'^4} \int d\vr_A d\vr_A^0 d\vr_{B_i} d\vr_{B_i}^0 H_{A,C_k} H_{B_i,C_k} H_{A^0,C_l} H_{B_i^0,C_l}\\
&& \times G_A(\vr_A,t;\vr_A^0,t^0) G_B(\vr_{B_i},t;\vr_{B_i}^0,t^0). \ee
Note that when $k=l$, $H_{A,C_k} H_{B_i,C_k} H_{A^0,C_l} H_{B_i^0,C_l}$ reduces to $H_{A,C_k} H_{B_i,C_k}$.

Finally, using the latter decomposition of the integral $I$, and
once again splitting it with respect to cases $k=l$ and $k\neq l$,
summing over $i,j,k,l$ we obtain:

\be \label{Iint} \nonumber I&=&m(m-1)qI_{ijkk;i\neq
j}+mqI_{iikk}+m(m-1)\\
&&\times q(q-1)I_{ijkl;i\neq j,k\neq l}+mq(q-1)I_{iikl;k\neq l}. \ee

Now we have to average the integral $I$ over the positions of the
catalytic subvolumens $C$. Performing such averaging and denoting it
by angle brackets with the subscript $\vr_C$, we have the following
four terms:

\be \nonumber \mbox{\textbf{(1) }}& \langle I_{ijkk;i\neq j}\rangle_{\vr_{C_k}}&=\frac{V^{m-2}}{V'^2}\int d\vr_A d\vr_A^0 H_{A} H_{A^0}
G_A(\vr_A,t;\vr_A^0,t^0) \\ \nonumber &&\equiv V^{m-2}\chi(t-t^0) \\ \mbox{\textbf{(2) }} \nonumber &\langle I_{ijkl;i\neq j,k\neq
l}\rangle_{\vr_{C_k},\vr_{C_l}}&=\frac{V^{m-2}}{V'^2}\frac{VV'^2}{V^2}=\frac{V^{m-2}}{V}
\\ \mbox{\textbf{(3) }} \nonumber &\langle
I_{iikk}\rangle_{\vr_{C_k}}&= \frac{V^{m-1}}{V'^4}\int d\vr_A d\vr_A^0 d\vr_{B_i} d\vr_{B_i}^0 \\ \nonumber &\times H_{A} H_{B_i}H_{A^0}
H_{B_i^0}\times & G_A(\vr_A,t;\vr_A^0,t^0)
G_B(\vr_{B_i},t;\vr_{B_i}^0,t^0) \\ \nonumber &&\equiv \frac{V^{m-1}}{V'^2}\kappa_1(t-t^0) \\
\mbox{\textbf{(4) }} \nonumber &\langle I_{iikl;k\neq l}\rangle_{\vr_{C_k},\vr_{C_l}}&= \frac{V^{m-1}}{V'^4}\frac{1}{V}\int d\vu d\vr_A d\vr_A^0
d\vr_{B_i}d\vr_{B_i}^0 \\ \nonumber &\times H_{A} H_{B_i}H_{A^0} H_{B_i^0}&\times G_A(\vr_A,t;\vr_A^0+\vu,t^0)
G_B(\vr_{B_i},t;\vr_{B_i}^0+\vu,t^0) \\
&&\equiv\frac{V^{m-1}}{VV'^2}\kappa_2(t-t^0) \ee where $\chi(t)$,
$\kappa_1(t)$ and $\kappa_2(t)$ are some functionals of time, which
will be made explicit in the next section.

Using the latter results, we may express the $v(t)$ function in
terms of the functionals $\chi(t-t^0)$, $\kappa_1(t-t^0)$ and
$\kappa_2(t-t^0)$. This gives

\be \label{vtF} \nonumber v(t)&=&v_{eq}-k\frac{(m-1)(q-1)}{V^2}\int_0^t dt^0 v(t^0)-k\frac{m-1}{V}\int_0^t dt^0v(t^0)\chi(t-t^0)\\
&&-\frac{k}{V'^2} \int_0^t dt^0v(t^0)\kappa_1(t-t^0)-k\frac{q-1}{VV'^2}\int_0^t dt^0 v(t^0)\kappa_2(t-t^0). \ee

Now, defining the Laplace transformation over the time variable of a
function $f(t)$ as $\wh{f}(s)\equiv \int_0^{\infty}dtf(t)e^{-st}$,
and performing the Laplace transformation of both sides of
Eq.(\ref{vtF}), we get:

\be \wh{v}(s)=\frac{v_{eq}}{s}-\frac{k v_{eq} \wh{v}(s)}{s}-k[B]\wh{v}(s)\wh{\chi}(s)-\frac{k}{V'^2}\wh{v}(s)\wh{\kappa}_1(s)
-\frac{k[C]}{V'^2}\wh{v}(s)\wh{\kappa}_2(s) \ee
where $\ds \frac{(m-1)(q-1)}{V^2}\simeq [B][C]=v_{eq}$, which implies that the Laplace-transformed $v(t)$ function is given explicitly by

\be \wh{v}(s)=\frac{v_{eq}}{s\left(1+k[B]\wh{\chi}(s)+\frac{k\wh{\kappa}_1(s)}{V'^2}+\frac{k[C]\wh{\kappa}_2(s)}{V'^2}+k v_{eq}\right)}. \ee

Now, in virtue of equation (\ref{relphiv}), $\wh{v}(s)$ and the
Laplace-transformed survival probability are related to each other
as

\be \wh{P}(s)=\frac{1-k\wh{v}(s)}{s} \ee

Hence, the Laplace-transformed survival probability obeys:

\be \label{pp}  \wh{P}(s)=\left[ s+\frac{k
v_{eq}}{1+k[B]\wh{\chi}(s)+\frac{k}{V'^2}\wh{\kappa}_1(s)+\frac{k[C]}{V'^2}\wh{\kappa}_2(s)}
\right]^{-1}. \ee It can be shown \cite{rice,wilemski73} that $\wh{\chi}(s)$, $\wh{\kappa}_1(s)$ and $\wh{\kappa}_2(s)$ can be replaced by their value for $s=0$ to obtain the long time behaviour of $P(t)$. Thus, inverting the latter equation in the limit $t\to \infty$, the $A$
particle survival probability $P(t)$ decays exponentially as:

\be P(t) \simeq
\exp{\left(-\frac{k[B][C]t}{1+k[B]\wh{\chi}(0)+\frac{k\wh{\kappa}_1(0)}{V'^2}+\frac{k[C]\wh{\kappa}_2(0)}{V'^2}}\right)}
\ee and consequently, comparing the latter equation against the
conventional form\be P(t) \simeq \exp{\left(- k_{eff} [B] t\right)}
\ee we have that the effective, overall reaction rate $k_{eff}$
describing the kinetics of diffusion-limited catalytically-activated
reactions obeys:

\be \label{lll}\frac{1}{k_{eff}} = \frac{1}{k [C]}+\frac{[B]
\wh{\chi}(0)}{[C]}+\frac{\wh{\kappa}_1(0)}{V'^2
[C]}+\frac{\wh{\kappa}_2(0)}{V'^2} \ee which can be thought of as
some "law of addition of inverse resistivities" and resembles
(although has a more complex form) the classical result of Collins
and Kimball for diffusion-limited trapping reactions with finite
elementary reaction constant $k$ \cite{collins}.

Equation (\ref{lll}) is the central result of our analysis. Functions $\wh{\chi}(0),\wh{\kappa}_1(0)$ and $\wh{\kappa}_2(0)$  are studied in the next section.


\section{Residence times}

In this section we show that functions
$\wh{\chi}(0),\wh{\kappa}_1(0)$ and $\wh{\kappa}_2(0)$ have an
apparent physical interpretation in terms of different residence
times of Brownian paths in finite domains and may be evaluated in
explicit form.

\subsection{One-particle's residence time}

The function $\wh{\chi}(0)$ entering the effective reaction rate is
defined by:

\be \wh{\chi}(0)=\frac{1}{V'^2}\int_0^{\infty}dt \int d\vr_A d\vr_A^0 H_{A} H_{A^0}G(\vr_A,t;\vr_A^0,0). \ee

\begin{figure}[htb]
\begin{center}
\includegraphics[scale=.6]{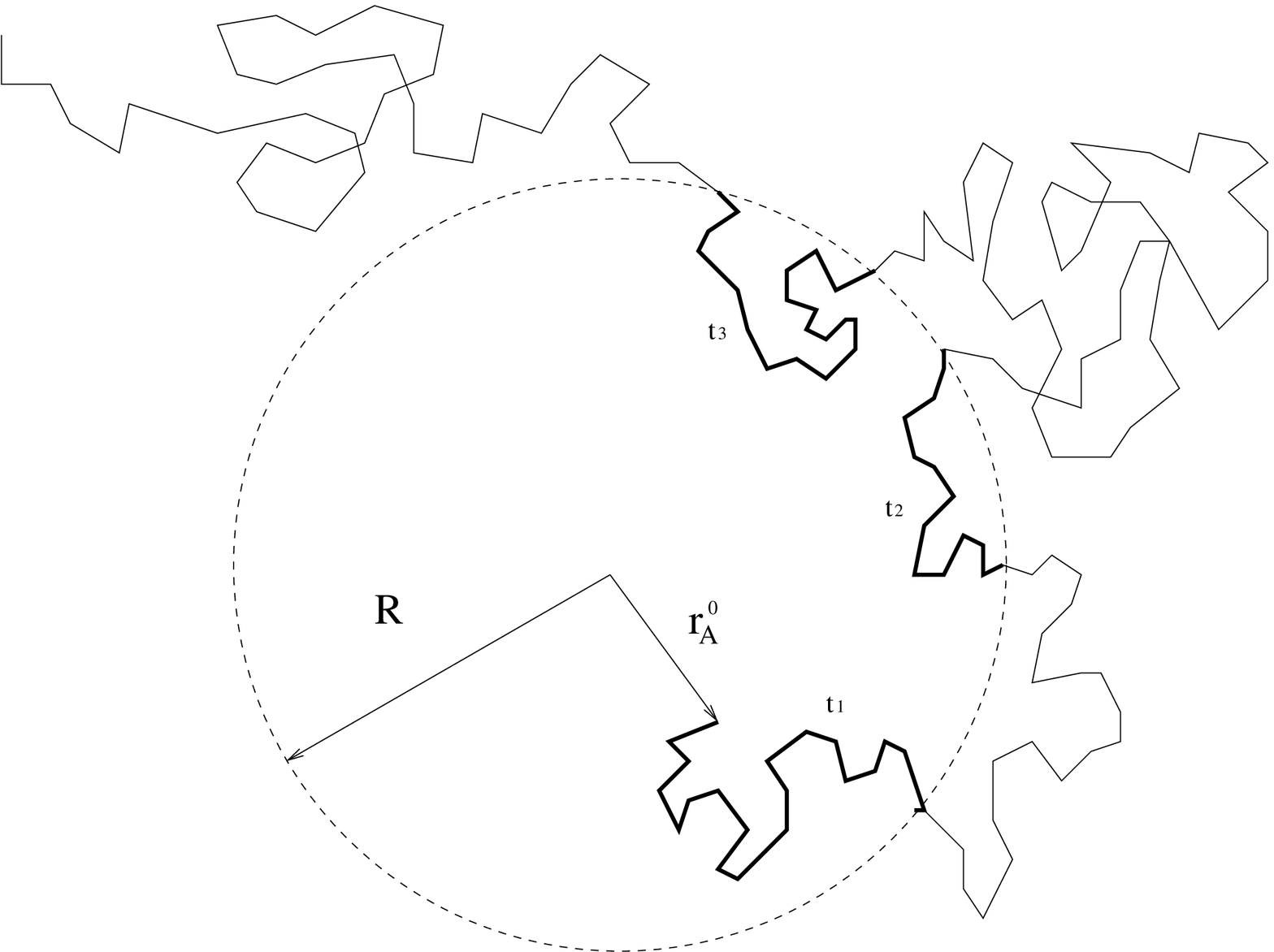}
\end{center}
\caption{\label{figtps2}}
\end{figure}

One notices now that $\wh{\chi}(0)$ can be interpreted as the total
time spent in a sphere of radius $R$ by a Brownian particle $A$,
which started its diffusion at time $t = 0$  at position $\vr_A^0$
(see figure \ref{figtps2}), averaged over all initial positions
inside this sphere. In other words, this time is the cumulative residence
time inside the sphere up to an infinite observation time. In one or
two dimensions it is infinite since the particle is certain to come
back to the sphere, but in three dimensions it is finite since the
particle can travel to infinity and thus is not certain to return.

This residence time is well known \cite{berezhkovskii98}, and can be
calculated rather straightforwardly. Indeed, integrating first over
the time with the change of variable $\ds y=\frac{1}{t}$, we obtain:

\be \wh{\chi}(0)=\frac{1}{V'^2}\int d\vr_A d\vr_A^0 H_{A}
H_{A^0}\frac{1}{4\pi D_A}\frac{1}{||\vr_A-\vr_A^0||} \ee Now, since
\be \ds
\int_0^{2\pi}\int_0^{\pi}\frac{1}{||\vr_A-\vr_A^0||}\sin{\theta}d\theta
d\phi=\frac{4\pi}{max(r_A,r_A^0)}, \ee we have:

\be \wh{\chi}(0)=\frac{1}{V'^2}\frac{4\pi}{D}\int_0^R dr_A r_A^2
\int_0^R dr_A^0 (r_A^0)^2 \frac{1}{max(r_A,r_A^0)}. \ee Splitting
next the second integral into the sum $\ds
\int_0^R=\int_0^{r_A}+\int_{r_A}^R$, we get: \be
\wh{\chi}(0)=\frac{1}{V'^2}\frac{16\pi R^5}{30D_A}, \ee which
yields, eventually, the following result:

\be \wh{\chi}(0)^{-1}=\frac{5}{6} 4\pi D_AR.  \ee We hasten to
remark that this expression, up to a numerical factor $5/6$,
coincides with the famous expression for the Smoluchowski reaction
constant in three dimensions, $K_S = 4\pi D_AR$.

\subsection{Two-particle's joint residence time}

\begin{figure}[htb]
\begin{center}
\includegraphics[scale=.6]{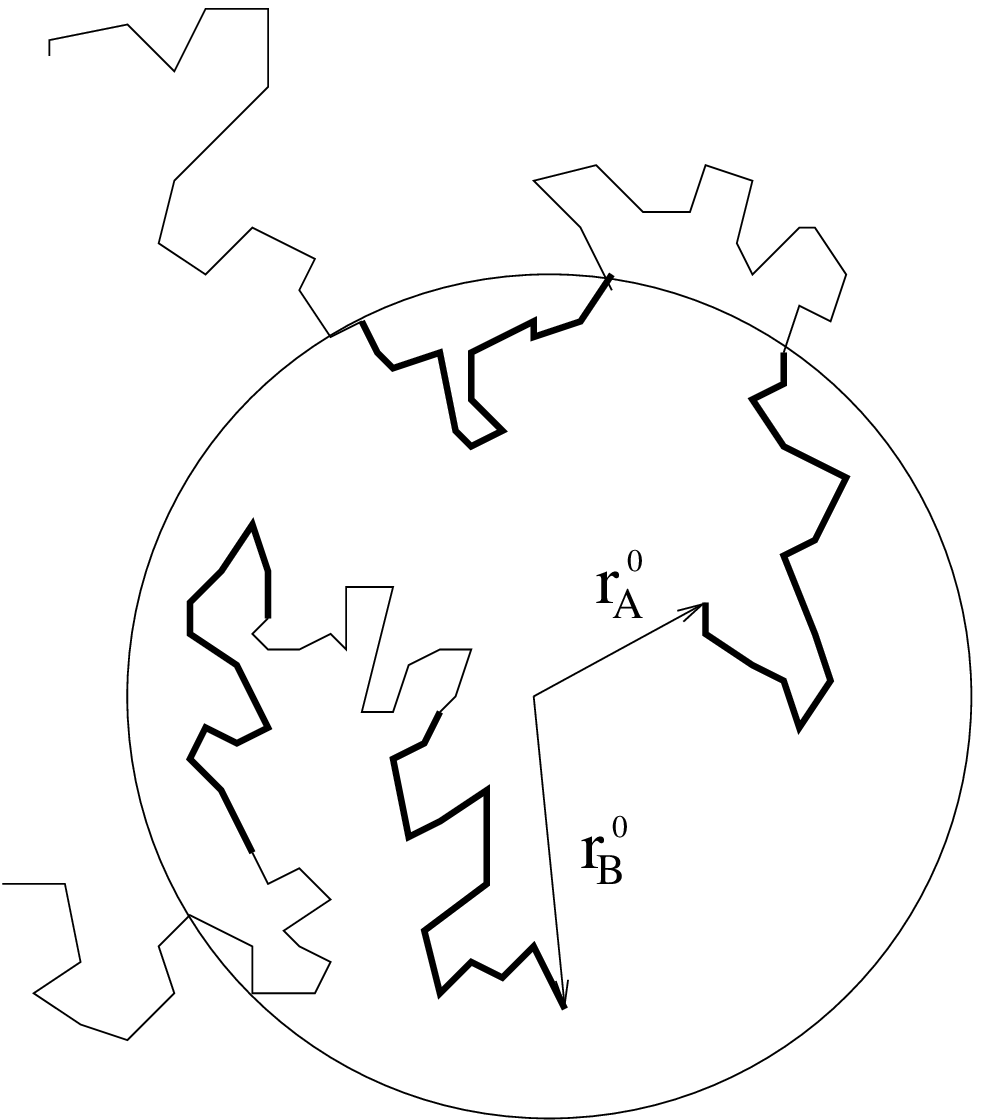}
\end{center}
\caption{\label{figtps3}}
\end{figure}

Now, we turn to two other functions - $\wh{\kappa}_1(0)$ and
$\wh{\kappa}_2(0)$, entering equation (\ref{ll}).  The first one is
formally defined by:

\be \label{kappa1} \nonumber
\wh{\kappa}_1(0)&=&\frac{1}{V'^2}\int_0^{\infty}dt\int d\vr_A d\vr_A^0 d\vr_{B_i} d\vr_{B_i}^0 H_{A} H_{B_i}\\
&& \times H_{A^0} H_{B_i^0} G(\vr_A,t;\vr_A^0,0)
G(\vr_{B_i},t;\vr_{B_i}^0,0).  \ee One may now notice that
$\wh{\kappa}_1(0)$ corresponds to the joint residence time in a
sphere of radius $R$ of particles $A$ and $B$, averaged over all
initial positions of particles $A$ and $B$ inside the sphere (see
figure \ref{figtps3}).

Such a residence time has been amply discussed  in another work
\cite{nousresidence}, and here we will merely present the result of
these calculations. It has been shown in Ref.\cite{nousresidence}
that $\wh{\kappa}_1(0)$ is a complicated function of the diffusion
coefficients $D_A$ and $D_B$ and is given explicitly by:

\be
\wh{\kappa}_1(0)=\frac{R^2}{20\pi}\left\{\frac{1}{D_A}m\left(\frac{D_A}{D_B}\right)+
\frac{1}{D_B}m\left(\frac{D_B}{D_A}\right)\right\} \ee where

\be \label{mx} m(x)=\frac{2-10\ln(1+x)}{x^{1/2}}-2\frac{\ln(1+x)}{x^{3/2}}+16\arctan(\sqrt{x})-\frac{7}{2} x^{1/2}.  \ee

Finally, the last undetermined function $\wh{\kappa}_2(0)$ obeys:

\be \nonumber \label{twojoint} \wh{\kappa}_2(0)=\frac{1}{V'^2}\int_0^{\infty}dt\int d\vu d\vr_A d\vr_A^0 d\vr_{B_i}d\vr_{B_i}^0 H_{A} H_{B_i}H_{A^0} H_{B_i^0} \\
\times G(\vr_A,t;\vr_A^0+\vu,0) G(\vr_{B_i},t;\vr_{B_i}^0+\vu,0). \ee

\begin{figure}[htb]
\begin{center}
\includegraphics[scale=.6]{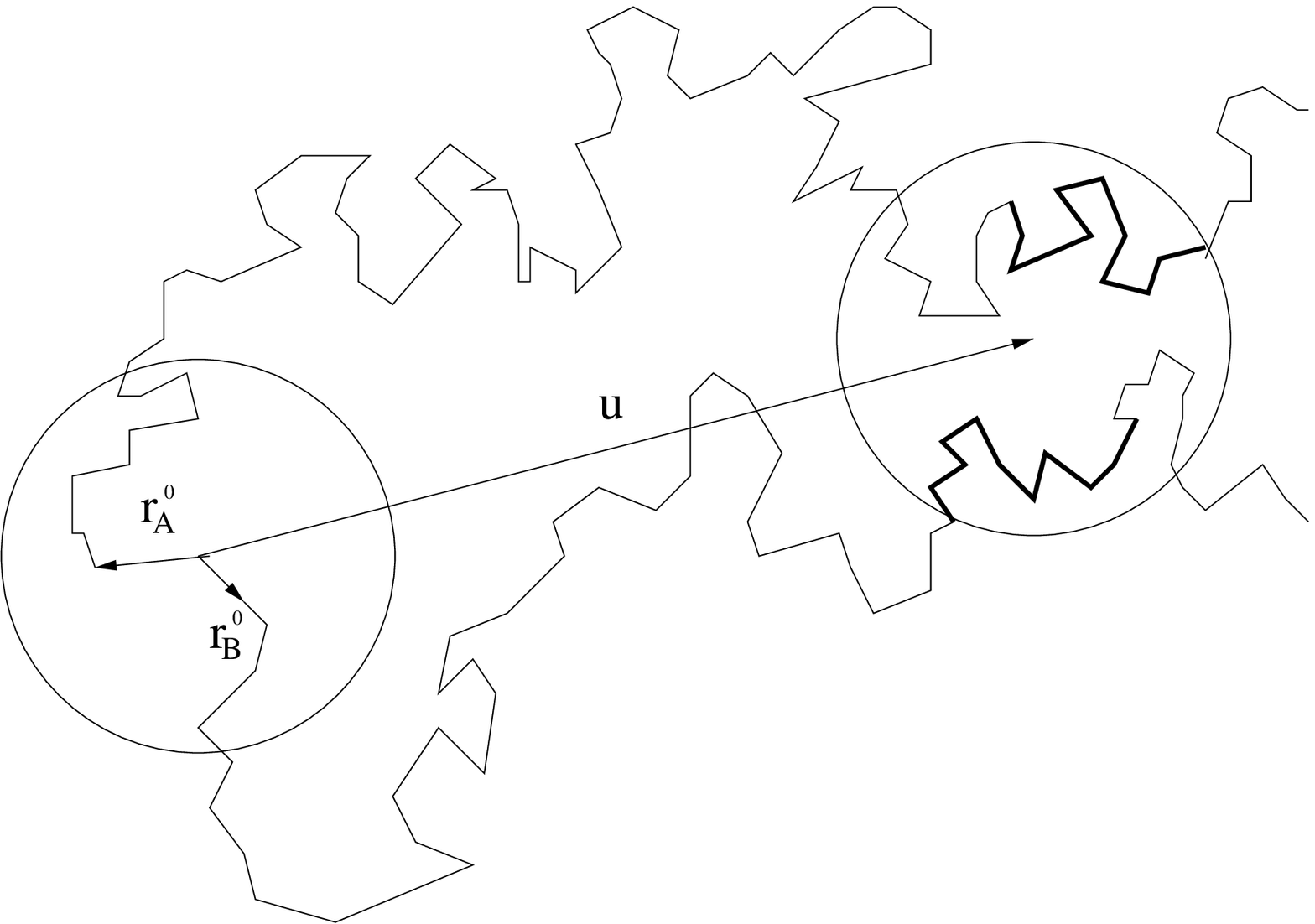}
\end{center}
\caption{\label{figtps4}}
\end{figure}

One notices that it may be interpreted as  the joint
residence time of particles $A$ and $B$ inside a sphere when the
particles initially start from a sphere separated by the vector
$\vu$ from the residence sphere (see figure \ref{figtps4}). Note
that the joint residence time $\wh{\kappa}_2(0)$ is obtained when
summing over all $\vu$.

It is shown in Appendix that:
\be \frac{V'^2}{\wh{\kappa}_2(0)}&=&4\pi R(D_A+D_B)\frac{385}{334}\approx 1.15\times 4 \pi R(D_A+D_B). \ee

\subsection{Effective reaction rate}

Eventually, summing up the results of this section, we present an
explicit expression for the effective reaction rate describing the
kinetics of diffusion-limited catalytically-activated trapping
reactions in terms of a suitably extended Wilemski and Fixman
approach:

\be \label{ll} \nonumber \frac{1}{k_{eff}} = \frac{1}{k [C]}
+\frac{[B]}{\frac{5}{6} 4\pi D_A R [C]} + \\
+\frac{1}{g(D_A,D_B)\pi^3R^4[C]} +\frac{1}{\frac{385}{334} 4 \pi
R(D_A+D_B)} \ee

where
$g^{-1}(D_A,D_B)=\frac{9}{320}\left\{\frac{1}{D_A}m\left(\frac{D_A}{D_B}\right)+
\frac{1}{D_B}m\left(\frac{D_B}{D_A}\right)\right\}$, $m(x)$ being
defined in equation (\ref{mx}). Note that the correlations that are
neglected in the Wilemski-Fixman approximation cannot be estimated
precisely, so that it is difficult to find how they affect the
different terms included into the reaction constant. However, these
terms should not be considered as successive corrections to the
usual expression for a bimolecular reaction. In fact, the joint
residence time of two molecules on a catalytic site cannot be
treated as a perturbation of the residence time of one molecule on
this site, except in some limit cases. Clearly, our results should
be compared to experimental or numerical results in order to be
discussed.

Consider now the behavior of the effective reaction constant in
Eq.(\ref{ll}) in several limiting cases. In non-catalytic systems,
in which the $A$ and $B$ particles may react at any point, which
corresponds to an evident situation with $[C] \to \infty$ (but $k
[C] = K$ is kept finite), we find from Eq.(\ref{ll}) that

\be \frac{1}{k_{eff}}\simeq \frac{1}{K} + \frac{1}{\frac{385}{334} 4
\pi (D_A + D_B) R}, \ee which represents,  up to a numerical factor
$385/334 \approx 1.15$ in the second term, the classical Collins and
Kimball result \cite{collins} describing the effective reaction rate
for trapping reactions involving diffusive $A$ and $B$ particles in
non-catalytic systems. Note that the numerical factor $385/334$
comes from the description of the reactive process which differs
between the Collins-Kimball approach and the present case.

Next, we turn to a different trivial situation when the particles $B$
are present in a great excess, i.e. $[B] \to \infty$, such that
their diffusion becomes irrelevant. In this limiting case we find
from Eq.(\ref{ll}) the following result: \be k_{eff} \simeq
\frac{5}{6} \frac{4 \pi D_A R [C]}{[B]}, \ee such that the $A$
particle survival probability follows \be P(t) \simeq
\exp{\left(-\frac{5}{6} 4 \pi D_A R [C] t \right)} \ee This is,
again, a standard Smoluchowski-type (up to a numerical factor $5/6$)
prediction for trapping $A+C\to C$ reactions with immobile traps C.

Finally, some simple analysis shows that for sufficiently small
$D_B$ the third term on the right-hand-side of Eq.(\ref{ll})
dominates and the effective reaction constant follows \be k_{eff}
\simeq g(D_A,D_B) \pi^3 R^4 [C] \ee Surprising feature of this
result is that $k_{eff}$ is proportional not to the first power of
the reaction radius $R$, but to the \textit{fourth} power of it!
Curiously enough, this prediction coincides with earlier results
obtained for trimolecular reaction of the form $A+A+C\to P+C$ using
an extended Collins-Kimball approach \cite{oshanin98}. This
anomalous dependence has been confirmed by Molecular Dynamics
simulations in Ref.\cite{tox}.

\section{Fluctuation-induced long time behavior}

As shown in the introduction for trapping reactions involving diffusive particles, some fluctuation states can change significantly the kinetics of diffusion-limited, catalytically-activated trapping
reactions. The particular systems with random placement of the catalytic
subvolumens the long-time kinetic behavior is
described by a stretched-exponential function, so that the usual kinetic laws do not hold.

Suppose that the $A$ particle is initially at the origin, the
traps $B$ are also initially uniformly spread in the reaction bath
and the immobile catalytic subvolumens $C$ are randomly distributed
in the reaction bath such that the \textit{closest} to the origin
(i.e. to the $A$ particle) subvolumen is at the distance $\rho$
apart from it. Then, the $A$ particle survival probability is
evidently bounded from below by \be \label{bou} P(t) \geq P_{lac}
\times P_{A}(\rho,t), \ee where $P_{lac}$ is the probability of
having a spherical region of radius $\rho$ completely devoid of the
\textit{catalytic subvolumens $C$} and $P_{A}(\rho,t)$ is, again,
the probability that the $A$ particle will not leave this region up
to time $t$; in these conditions, the diffusive $A$ particle may
meet the diffusive traps $B$ many times but the reaction can not
take place since the necessary ingredient of the elementary act -
the presence of the catalytic subvolumen $C$, will not be fulfilled.

Consequently, the $A$ particle survival probability obeys \be P(t)
\geq \exp{\left(- \frac{4}{3} \pi \rho^3 [C] \right)} \times
\exp{\left(- \frac{D_A}{\rho^2} t \right)} \ee This bound is
valid for any $\rho$ and we have to choose such value of it, which
provides the maximal lower bound. Maximizing the right-hand-side of
the latter equation, we find that the optimal $\rho$ is $\rho \sim
(D_A t/2 \pi [C])^{1/5}$, which yields \be \label{too} P(t) \geq
\exp{\left(- \frac{3}{5} (2 \pi)^{2/5} [C]^{2/5} (D_A
t)^{3/5}\right)}, \ee i.e. the law similar to the one in
Eq.(\ref{traps}) describing the long-time evolution of the survival
probability of a particle diffusing in presence of immobile,
randomly placed traps.

Since the right-hand-side of Eq.(\ref{too}) decays slower than
exponentially, we may infer that at long-times the kinetics of
diffusion-limited catalytically-activated trapping reactions will be
supported by such fluctuation states and will be described by a
stretched-exponential function of time. The comparison of \ref{too} with the classical kinetic law given by \ref{ll} shows that in all conditions the bound \ref{too} should only be considered for excecdingly long times, so that the classical law usually holds.

\section{Conclusion}

To conclude, in this paper we have studied the kinetics of the
catalytically-activated diffusion-limited reactions in
Eq.(\ref{rea}) in the special case when $B$ particles remain
unaltered by reactions, i.e. the case of diffusion-limited
catalytically-activated trapping reactions. In order to obtain an
effective reaction rate for such bi-catalytic reactions, we have
developed an analytical approach based on the work of Wilemski and
Fixman \cite{wilemski73}, which allowed us to calculate analytically
the effective reaction constant.  We have shown that this effective
reaction constant comprises several terms which may be interpreted
in terms of the residence and joint residence times of Brownian
paths in finite domains. We have demonstrated that the effective
reaction constant exhibits a non-trivial dependence on the reaction
radii, the mean density of catalytic subvolumens and particles'
diffusion coefficients. Finally, we have discussed  the impact of
several fluctuation states giving rise to anomalous
fluctuation-induced contributions to the long-time kinetic behavior in such systems. Except in this asymptotic case, however the usual kinetic laws hold with the effective reaction constant calculated previously. These results can be very useful not only in the theory of heterogeneous catalysis, but also in biology, when reactions can only take place on specific sites.

\section{Acknowledgments}

The authors wish to thank Prof. M.Tachiya and Prof. H.Wio for
fruitful discussions. G.O. acknowledges the financial support from
the Alexander von Humboldt Foundation via the Bessel Research Award.

\section{Appendix: Joint residence time in a distant sphere}

In order to evaluate explicitly the function given in \ref{twojoint}, we first integrate the
two propagators over $\vu$:

\be \nonumber \int d\vu G(\vr_A+\vu,t|\vr_A^0,0)
G(\vr_{B}+\vu,t|\vr_{B}^0,0)\\
\nonumber =\int d\vX G_{\frac{D_AD_B}{D_A+D_B}}(\vX,t|\vX^0,0) G_{D_A+D_B}(\vr_{B}-\vr_A,t|\vr_{B}^0-\vr_A^0,0) \\
= G_{D_A+D_B}(\vr_{B}-\vr_A,t|\vr_{B}^0-\vr_A^0,0), \ee where

\be \nonumber
\vX&=&\frac{D_B}{D_A+D_B}(\vr_A+\vu)+\frac{D_A}{D_A+D_B}(\vr_B+\vu)\\
\vX^0&=&\frac{D_B}{D_A+D_B}(\vr_A^0+\vu)+\frac{D_A}{D_A+D_B}(\vr_B^0+\vu)
\ee and $G_D$ is the Gaussian propagator associated to the diffusion
coefficient $D$.

\begin{figure}[htb]
\begin{center}
\includegraphics*[scale=0.4]{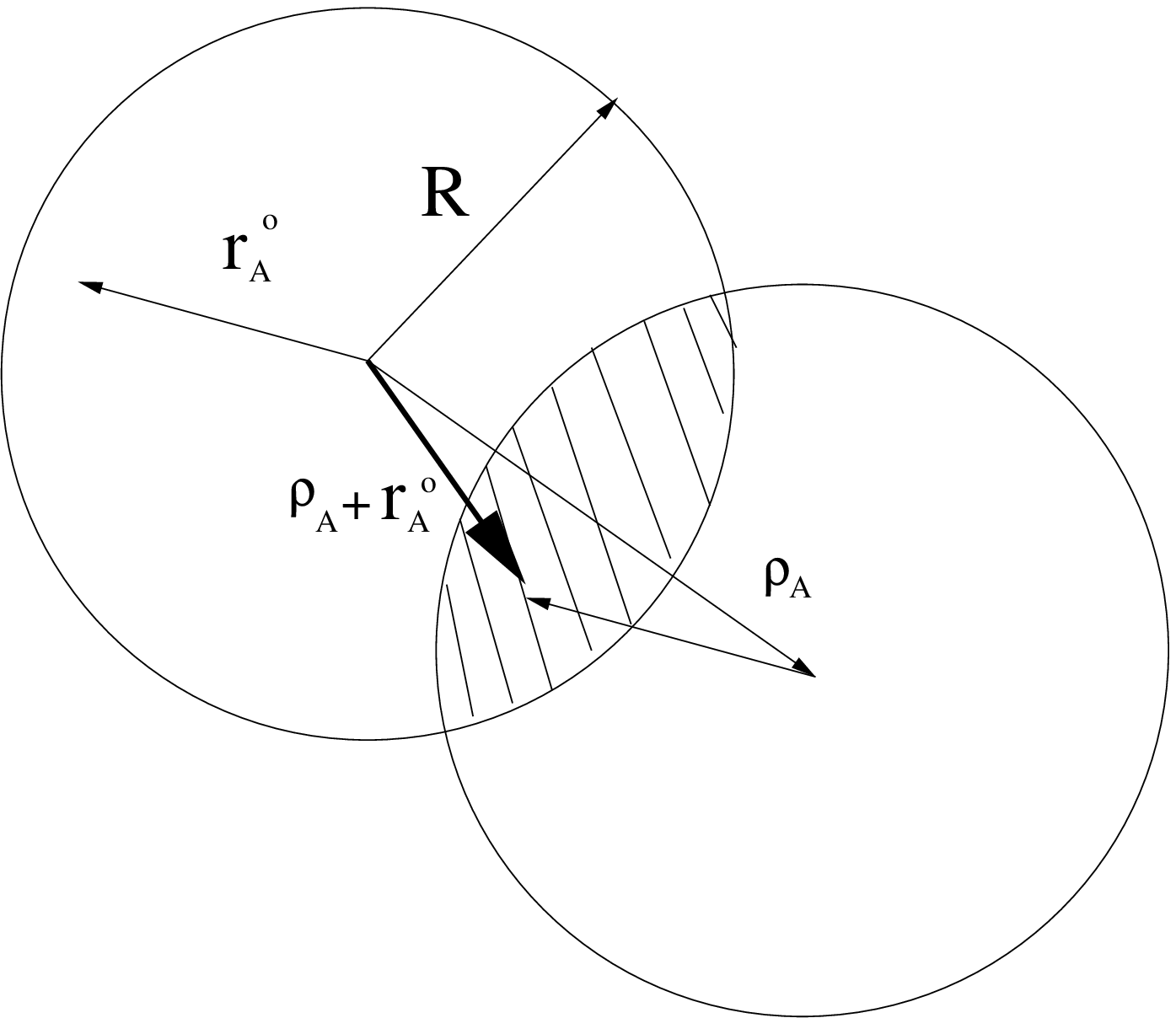}
\caption{\label{recouvs}}
\end{center}
\end{figure}

Then, the integral defined in (\ref{twojoint}) attains the following
form:

\be \nonumber V'^2\wh{\kappa}_2(0)=\int_0^{\infty}dtd\vr_A^0 d\vr_{B}^0 d\vro_A
d\vro_B H_{A^0} H_{B^0}\\
H_{\vro_A,A^0} H{\vro_B,B^0} G_{D_A+D_B}(\vro_{B}-\vro_A,t|\vze,0)
\ee where $\vro_B=\vr_B-\vr_B^0$ and $\vro_A=\vr_A-\vr_A^0$. The
calculation of this equation requires the analysis of integrals of
the following type:

\be \int d\vr_A^0H_{A^0}H_{\vro_A,A^0} \ee which represent the
overlapping area between two spheres (see figure \ref{recouvs}).

This area can be straightforwardly obtained:

\be \mathcal{A}&=&2\int_{\rho_A/2}^{R}dr \pi (R^2-r^2)=\frac{\pi}{3}(R-\rho_A/2)^2(4R+\rho_A). \ee

Next, we have to calculate the integral

\be \nonumber V'^2\wh{\kappa}_2(0)=\int_0^{\infty}dtd\vro_A d\vro_B
H(2R-|\vro_A|)H(2R-|\vro_B|)\\
\nonumber
\left[\frac{\pi}{3}(R-\vro_A/2)^2(4R+\vro_A)\right]\left[\frac{\pi}{3}(R-\vro_B/2)^2(4R+\vro_B)\right]\\
\times G_{D_A+D_B}(\vro_{B}-\vro_A,t|\vze,0). \ee Using the explicit
expression for the propagators and integrating them over the time
variable, we obtain

\be I=\int d\vro_A d\vro_B H(2R-|\vro_A|)H(2R-|\vro_B|)
\frac{\rho_A^n\rho_B^m}{||\vro_B-\vro_A||} \ee and

\be
\int_0^{2\pi}\int_0^{\pi}\frac{1}{||\vro_B-\vro_A||}\sin{\theta_B}d\theta_Bd\phi_B=\frac{4\pi}{max(\rho_A,\rho_B)}
\ee which yields

\be I=(4\pi)^2\int_0^{2R}d\rho_A \rho_A^{n+2}\int_0^{2R}d\rho_B
\rho_B^{m+2}\frac{1}{max(\rho_A,\rho_B)} \ee Consequently, the
latter equation enables us to evaluate an
 explicit expression for $\wh{\kappa}_2(0)$:

\be \frac{V'^2}{\wh{\kappa}_2(0)}&=&4\pi R(D_A+D_B)\frac{385}{334}\approx 1.15\times 4 \pi R(D_A+D_B). \ee

\newpage

\newpage

\newpage

\newpage

\newpage

\newpage

\newpage


{\large \textbf{Figure legend}}

\bigskip

\textbf{Figure 1.} Schematic representation of the bi-catalytic reaction:
volume $V$, comprising a single  $A$ particle, $m$ diffusing $B$ particles and $q$ immobile
subvolumen $C$.

\bigskip

\textbf{Figure 2.} Trimolecular reaction: the trajectories in bold
type are those for which the reaction takes place, i.e. when one
particle $A$ and one $B$ are jointly present in a catalytic domain.

\bigskip

\textbf{Figure 3.} Residence time  of a Brownian particle in a sphere.

\bigskip

\textbf{Figure 4.} Joint residence time in a sphere. Blod lines: trajectories during common residence time.

\bigskip

\textbf{Figure 5.} Joint residence time in a distant sphere.

\bigskip

\textbf{Figure 6.} Overlapping area between two spheres.


\end{document}